\documentclass[english,aps,pra,superscriptaddress,notitlepage]{revtex4-1}
\usepackage[latin9]{inputenc}
\usepackage{color}
\usepackage{babel}
\usepackage{amsmath}
\usepackage{amssymb}
\usepackage[unicode=true,pdfusetitle,
 bookmarks=true,bookmarksnumbered=false,bookmarksopen=false,
 breaklinks=false,pdfborder={0 0 1},backref=false,colorlinks=true]
 {hyperref}
\hypersetup{
 urlcolor=blue, citecolor=blue, hyperfootnotes=blue, linkcolor=blue}
\usepackage{breakurl}

\makeatletter
\usepackage{mathrsfs}
\usepackage{epsfig}
\usepackage{amsfonts}
\usepackage[figuresright]{rotating}
\usepackage{dcolumn}
\usepackage{bm}
\usepackage{color}

\makeatother

\begin{document}

\title{Cooling microwave fields into general multimode Gaussian states}

\author{Nahid Yazdi}

\affiliation{Department of Physics, Isfahan University of Technology, Isfahan
84156-83111, Iran}

\affiliation{Wilczek Quantum Center, School of Physics and Astronomy, Shanghai
Jiao Tong University, Shanghai 200240, China}

\author{Juan José García-Ripoll}

\affiliation{Institute of Fundamental Physics IFF-CSIC, Calle Serrano 113b, 28006
Madrid, Spain.}

\author{Diego Porras}

\affiliation{Institute of Fundamental Physics IFF-CSIC, Calle Serrano 113b, 28006
Madrid, Spain.}

\author{Carlos Navarrete-Benlloch}
\email{corresponding author; derekkorg@gmail.com}

\selectlanguage{english}%

\affiliation{Wilczek Quantum Center, School of Physics and Astronomy, Shanghai
Jiao Tong University, Shanghai 200240, China}

\affiliation{Max-Planck Institute for the Science of Light, Staudtstrasse 2, 91058
Erlangen, Germany}

\affiliation{Shanghai Research Center for Quantum Sciences, Shanghai 201315, China}
\begin{abstract}
We show that a collection of lossy multi-chromatically modulated qubits
can be used to dissipatively engineer arbitrary Gaussian states of
a set of bosonic modes. Our ideas are especially suited to superconducting-circuit
architectures, where all the required ingredients are experimentally
available. The generation of such multimode Gaussian states is necessary
for many applications, most notably measurement-based quantum computation.
We build upon some of our previous proposals, where we showed how
to generate single-mode and two-mode squeezed states through cooling
and lasing. Special care must be taken when extending these ideas
to many bosonic modes, and we discuss here how to overcome all the
limitations and hurdles that naturally appear. We illustrate our ideas
with a fully worked out example consisting of GHZ states, but have
also tested several other examples such as cluster states. All these
examples allow us to show that it is possible to use a set of $N$
lossy qubits to cool down a bosonic chain of $N$ modes to any desired
Gaussian state.
\end{abstract}
\maketitle

\section{Introduction}

The generation of complex quantum states of many optical modes has
been on the roadmap of quantum optics for quite some time \citep{Fabre20}.
Apart from their fundamental motivation on questions of entanglement
\citep{Gaussian2}, such states are necessary for technological applications
such as measurement-based quantum computation \citep{MBQC1,MBQC2,MBQC3}.
While tremendous developments have been possible in this area thanks
to nonlinear optical cavities \citep{Wang20,Fabre15,Pfister14,Bachor12},
the generation of such states remains very challenging in the optical
domain. On the other hand, the development of quantum microwave physics
in the form of superconducting circuits \citep{JJGRbook,Blais21,Devoret13}
has allowed us to access regimes that remained largely unexplored
with other experimental platforms, for example, the ultra-strong coupling
regime of light-matter interactions \citep{USC1,USC2,USC3,USC4}.
This is mainly due to the low characteristic frequencies of these
systems, on the GHz domain, together with the large effective dipole
moments of the structures, reason why they are sometimes dubbed ``giant''
artificial atoms.

Exploiting the low characteristic frequencies of these systems, in
this work we put forward a proposal for the generation of general
multimode Gaussian states of microwave fields. Our idea relies on
the ability to modulate parameters of superconducting circuits at
rates comparable to their natural energy scales \citep{Li13,Lahteenmaki13}.
In fact, in previous works we have used similar ideas to show that
squeezed states of microwave fields can be generated through cooling
\citep{Porras12} or lasing \citep{CNB14}, but restricted there to
single-mode or two-mode Gaussian states. In the present work, we examine
the possibility of using similar ideas to generate arbitrary Gaussian
states of as many modes as one wants. We provide a positive answer,
but not without several subtleties that impose nontrivial conditions
that are necessary to examine in detail. From a more general point
of view, we study how to modulate a set of lossy qubits to dissipatively
engineer arbitrary Gaussian states of a bosonic chain.

The article is structured as follows. In the next section we introduce
the characterization of multimode Gaussian states, and present the
core of our idea that uses lossy qubits to cool down the bosonic modes
to the desired Gaussian state. In Section \ref{Sec:FormalIdea} we
develop the idea formally and discuss some potential limitations that
were not present for single-mode or two-mode states. In Section \ref{Sec:Example}
we work out a detailed example, the generation of Greenberger-Horne-Zeilinger
(GHZ) states; we first introduce them, to then show how to obtain
them with our ideas, finally showing that the limitations mentioned
in the previous section do not spoil our proposal. Throughout the
article, the idea is presented via a model in which all qubits are
coupled to all bosonic modes; since this might be highly impractical,
in Section \ref{Sec:AvoidingALLtoALL} we propose alternative models
with local couplings only, that allows implementing our ideas as well.
We finish the article in Section \ref{Sec:Conclusions} where we offer
some conclusions and comment on how to extend the idea to generate
multimode nonclassical lasing.

\section{Multimode Gaussian states and introduction to the generic idea\label{Sec:GaussianStatesIdea}}

\subsection{Characterization of Gaussian states\label{Sec:GaussianStates}}

Let us first establish what we mean by general Gaussian states \citep{CNBbook,Gaussian1,Gaussian2,ParisBook,CNB-QOnotes}.
Consider for this $N$ bosonic modes with annihilation operators that
we collect into the vector $\hat{\boldsymbol{a}}=(\hat{a}_{1},\hat{a}_{2},...,\hat{a}_{N})^{T}$,
satisfying canonical commutation relations $[\hat{a}_{j},\hat{a}_{l}^{\dagger}]=\delta_{jl}$
and $[\hat{a}_{j},\hat{a}_{l}]=0$. Any Gaussian state (up to a trivial
displacement) can be generated by applying a Gaussian unitary $\hat{G}$
to a thermal state of all modes \citep{CNBbook,Gaussian1,Gaussian2,ParisBook,CNB-QOnotes}
\begin{equation}
\hat{\rho}_{G}(\bar{\boldsymbol{n}})=\hat{G}\hat{\rho}_{\text{th}}(\bar{\boldsymbol{n}})\hat{G}^{\dagger}.
\end{equation}
with $\hat{\rho}_{\text{th}}(\bar{\boldsymbol{n}})=\otimes_{j=1}^{N}\hat{\rho}_{\text{th},j}(\bar{n}_{j})$,
where $\bar{\boldsymbol{n}}=(\bar{n}_{1},...,\bar{n}_{N})$ collects
the number of thermal excitations of the modes, which in turn fix
the entropy or mixedness of the state of the system, and
\begin{equation}
\hat{\rho}_{\text{th},j}(\bar{n}_{j})=\frac{e^{-\kappa_{j}\hat{a}_{j}^{\dagger}\hat{a}_{j}}}{\text{tr}\left\{ e^{-\kappa_{j}\hat{a}_{j}^{\dagger}\hat{a}_{j}}\right\} },
\end{equation}
is a thermal state for a mode with normalized inverse temperature
$\kappa_{j}$, related to the number of excitations by the Bose-Einstein
distribution $\bar{n}_{j}=\left(e^{\kappa_{j}}-1\right)^{-1}$. The
Gaussian unitary does not add any extra entropy to the state, but
changes the correlations (including entanglement) between the modes.
Such unitaries are characterized by having a linear action onto the
modes (note that we use an economic notation in which the transpose
symbol only transposes the vectors but without affecting the operators
that made them up, whereas the dagger symbol affects also the internal
operators)
\begin{equation}
\hat{G}^{\dagger}\hat{\boldsymbol{a}}\hat{G}=\mathcal{A}\hat{\boldsymbol{a}}+\mathcal{B}\hat{\boldsymbol{a}}^{\dagger T}\equiv\boldsymbol{\hat{A}},\label{A}
\end{equation}
with $N\times N$ complex matrices $\mathcal{A}$ and $\mathcal{B}$
subject to the constraints
\begin{equation}
\mathcal{A}\mathcal{B}^{T}=\mathcal{B}\mathcal{A}^{T},\;\mathcal{A}\mathcal{A}^{\dagger}=\mathcal{B}\mathcal{B}^{\dagger}+\mathcal{I},\label{GaussianUnitaryConditions}
\end{equation}
where $\mathcal{I}$ is the $N\times N$ identity, such that the transformed
annihilation operators $\boldsymbol{\hat{A}}$ satisfy canonical commutation
relations just like the original ones.

Pure states, for which $\bar{\boldsymbol{n}}=0$, correspond to the
Gaussian unitary acting on the vacuum of the original modes,
\begin{equation}
|G\rangle=\hat{G}|vac\rangle_{a},\;\text{ where }\hat{\boldsymbol{a}}|vac\rangle_{a}=0.\label{Gpure}
\end{equation}
In turn, this state is nothing but the vacuum of the transformed modes,
that is, $|G\rangle=|vac\rangle_{A}$, where $\boldsymbol{\hat{A}}|vac\rangle_{A}=0$.

It is useful to know the relation between the covariance matrix of
the Gaussian state and matrices $\mathcal{A}$ and $\mathcal{B}$.
Defining the vector of quadratures $\hat{\boldsymbol{r}}=(\hat{x}_{1},...,\hat{x}_{N},\hat{p}_{1},...,\hat{p}_{N})^{T}$,
with $\hat{x}_{j}=\hat{a}_{j}+\hat{a}_{j}^{\dagger}$ and $\hat{p}_{j}=-\mathrm{i}(\hat{a}_{j}-\hat{a}_{j}^{\dagger})$,
the covariance matrix elements are defined as $V_{mn}=\langle\hat{r}_{m}\hat{r}_{n}+\hat{r}_{n}\hat{r}_{m}\rangle/2$.
Using the commutation relations $[\hat{r}_{m},\hat{r}_{n}]=2\mathrm{i}\Omega_{mn}$
and the relation $\hat{\boldsymbol{r}}=\mathcal{T}\hat{\boldsymbol{\alpha}}$,
with\begin{subequations}
\begin{align}
\hat{\boldsymbol{\alpha}} & =\left(\begin{array}{c}
\hat{\boldsymbol{a}}\\
\hat{\boldsymbol{a}}^{\dagger T}
\end{array}\right)=(\hat{a}_{1},...,\hat{a}_{N},\hat{a}_{1}^{\dagger},...,\hat{a}_{N}^{\dagger})^{T},\\
\mathcal{T} & =\left(\begin{array}{cc}
\mathcal{I} & \mathcal{I}\\
-\mathrm{i}\mathcal{I} & \mathrm{i}\mathcal{I}
\end{array}\right),\\
\Omega & =\left(\begin{array}{cc}
0 & \mathcal{I}\\
-\mathcal{I} & 0
\end{array}\right),
\end{align}
\end{subequations}the covariance matrix can be written as
\begin{equation}
V=\mathcal{T}\underbrace{\langle\hat{\boldsymbol{\alpha}}\hat{\boldsymbol{\alpha}}^{T}\rangle}_{C}\mathcal{T}^{T}-\mathrm{i}\Omega.\label{CovarianceMatrixGaussian}
\end{equation}
In turn, the complex covariance matrix $C$ can be easily found in
terms of $\mathcal{A}$ and $\mathcal{B}$ as
\begin{align}
C & =\text{tr}\left\{ \hat{\rho}_{G}(\bar{\boldsymbol{n}})\hat{\boldsymbol{\alpha}}\hat{\boldsymbol{\alpha}}^{T}\right\} =\text{tr}\left\{ \hat{\rho}_{\text{th}}(\bar{\boldsymbol{n}})\hat{G}^{\dagger}\hat{\boldsymbol{\alpha}}\hat{\boldsymbol{\alpha}}^{T}\hat{G}\right\} \nonumber \\
 & =\left(\begin{array}{cc}
\mathcal{J}(\mathcal{A},\mathcal{B},\mathcal{B},\mathcal{A}) & \mathcal{J}(\mathcal{A},\mathcal{A}^{*},\mathcal{B},\mathcal{B}^{*})\\
\mathcal{J}(\mathcal{B}^{*},\mathcal{B},\mathcal{A}^{*},\mathcal{A}) & \mathcal{J}(\mathcal{B}^{*},\mathcal{A}^{*},\mathcal{A}^{*},\mathcal{B}^{*})
\end{array}\right),
\end{align}
where $\mathcal{J}(\mathcal{X},\mathcal{Y},\mathcal{Z},\mathcal{W})=\mathcal{X}(\mathcal{I}+\bar{\mathcal{N}})\mathcal{Y}^{T}+\mathcal{Z}\bar{\mathcal{N}}\mathcal{W}^{T}$,
$\bar{\mathcal{N}}=\text{diag}(\bar{\boldsymbol{n}})$ is a diagonal
matrix containing all the thermal populations in the diagonal, and
we have used (\ref{A}), as well as $\text{tr}\left\{ \hat{\rho}_{\text{th}}(\bar{\boldsymbol{n}})\hat{\boldsymbol{a}}^{\dagger T}\hat{\boldsymbol{a}}^{T}\right\} =\mathcal{\bar{N}}$,
$\text{tr}\left\{ \hat{\rho}_{\text{th}}(\bar{\boldsymbol{n}})\hat{\boldsymbol{a}}^{\dagger T}\hat{\boldsymbol{a}}^{T}\right\} =\mathcal{I}+\mathcal{\bar{N}}$,
and $\text{tr}\left\{ \hat{\rho}_{\text{th}}(\bar{\boldsymbol{n}})\hat{\boldsymbol{a}}\hat{\boldsymbol{a}}^{T}\right\} =0=\text{tr}\left\{ \hat{\rho}_{\text{th}}(\bar{\boldsymbol{n}})\hat{\boldsymbol{a}}^{\dagger T}\hat{\boldsymbol{a}}^{\dagger}\right\} $.

In the following, and as we did in this section, indices $j$, $l$,
and $k$ will run from 1 to $N$, while index $m$ will run up to
$2N$.

\subsection{Basic idea for the dissipative generation of Gaussian states\label{Sec:Idea} }

Our strategy in order to generate the multimode Gaussian states introduced
above is similar to the one we introduced in previous works for single-mode
and two-mode squeezed states \citep{Porras12,CNB14}. We couple $N$
modes of linear superconducting circuits with distinct frequencies
$\{\omega_{j}\}_{j=1,2,...,N}$ to $N$ superconducting qubits also
with distinct frequencies $\{\varepsilon_{j}\}_{j=1,2,...,N}$, as
described by the Hamiltonian (in Section \ref{Sec:AvoidingALLtoALL}
we explain how to avoid all-to-all couplings, and implement the idea
with local couplings only)
\begin{equation}
\hat{H}(t)=\sum_{j=1}^{N}\left(\omega_{j}\hat{n}_{j}+\frac{\varepsilon_{j}}{2}\hat{\sigma}_{j}^{z}\right)+\sum_{jl=1}^{N}g_{jl}(\hat{\sigma}_{j}+\hat{\sigma}_{j}^{\dagger})(\hat{a}_{l}+\hat{a}_{l}^{\dagger})+\sum_{j=1}^{N}\left[\sum_{m=1}^{2N}\Omega_{jm}\eta_{jm}\cos(\Omega_{jm}t+\phi_{jm})\right]\hat{\sigma}_{j}^{z},\label{Horiginal}
\end{equation}
with number operators $\hat{n}_{j}=\hat{a}_{j}^{\dagger}\hat{a}_{j}$,
and Pauli operators $\hat{\sigma}_{j}^{z}=|e\rangle_{j}\langle e|-|g\rangle_{j}\langle g|$
and $\hat{\sigma}_{j}=|g\rangle_{j}\langle e|$ for qubit $j$ with
ground and excited states $|g\rangle_{j}$ and $|e\rangle_{j}$, respectively.
Note that we are using $\hbar$ units for the Hamiltonian, so that
all parameters have frequency units for convenience. We assume that
all direct processes are far off resonant, $|\varepsilon_{j}\pm\omega_{l}|\gg|g_{jl}|$,
and add a temporal modulation of the qubit frequencies which will
help the system bring certain processes to resonance effectively.
In particular, in general we will need to modulate each qubit with
$2N$ different frequencies $\Omega_{jm}$, with corresponding (normalized)
amplitudes $0<\eta_{jm}\ll1$ and phases $\phi_{jm}\in[0,2\pi[$,
in order to be able to tune all the possible couplings between the
modes and the qubits. Of course, for specific states the final count
might be smaller.

As we show explicitly below, choosing the modulation frequencies 
\begin{equation}
\Omega_{jl}=\varepsilon_{j}-\omega_{l},\;\;\Omega_{j,N+l}=\varepsilon_{j}+\omega_{l},\;\;j,l=1,2,...,N\label{ModulationFreqs}
\end{equation}
we will be able to control all couplings of the qubits' ladder operators
to the modes' annihilation and creation operators, generating the
effective Hamiltonian
\begin{equation}
\hat{H}_{\text{eff}}=-\sum_{j=1}^{N}\bar{g}_{j}\hat{A}_{j}\hat{\sigma}_{j}^{\dagger}+\text{H.c.},\label{Heff}
\end{equation}
where $\bar{g}_{j}$ are some effective couplings and $\hat{A}_{j}$
are the transformed annihilation operators (\ref{A}) corresponding
to the Gaussian state $|G\rangle$ that we want to generate. Note
that that $\Omega_{jl}$ is precisely the energy missing to bring
the term $\hat{\sigma}_{j}\hat{a}_{l}^{\dagger}$ (and its Hermitian
conjugate) to resonance; similarly $\Omega_{j,N+l}$ provides the
energy missing for the $\hat{\sigma}_{j}\hat{a}_{l}$ process to play
a role. It is then intuitive that, in the right picture and under
the right conditions, (\ref{Heff}) will capture the physics of the
dynamics generated by (\ref{Horiginal}). We will prove this rigorously
shortly.

The final step consists on introducing a strong radiative decay on
each qubit at rate $\gamma_{j}\gg|\bar{g}_{j}|$. Hence, every time
an excitation is transferred from modes $\boldsymbol{\hat{A}}$ to
the qubits via (\ref{Heff}), the excitation will be quickly lost
before it can come back to the photonic modes, which will then be
cooled down the their vacuum state $|vac\rangle_{A}$ at rate $|\bar{g}_{j}|^{2}/\gamma_{j}$
\citep{Porras12,CNB14}. In particular, eliminating adiabatically
the qubits using standard techniques \citep{Porras12,CNB14,CNB-QOnotes},
the reduced state $\hat{\rho}$ of the bosonic modes is easily shown
to obey the following master equation:
\begin{equation}
\frac{d\hat{\rho}}{dt}=\sum_{j=1}^{N}\left(\frac{|\bar{g}_{j}|^{2}}{\gamma_{j}}\mathcal{D}_{A_{j}}[\hat{\rho}]+\kappa\mathcal{D}_{a_{j}}[\hat{\rho}]\right),
\end{equation}
with $\mathcal{D}_{C}[\hat{\rho}]=2\hat{C}\hat{\rho}\hat{C}^{\dagger}-\hat{C}^{\dagger}\hat{C}\hat{\rho}-\hat{\rho}\hat{C}^{\dagger}\hat{C}$.
Note that we have taken into account the decay of the original modes
$\hat{a}_{j}$ at rates $\kappa$. In the limit of large cooperativities
$|\bar{g}_{j}|^{2}/\gamma_{j}\kappa\gg1$ the local decays $\mathcal{D}_{a_{j}}$
are negligible, so that the dominant $\mathcal{D}_{A_{j}}$ terms
will steer the state into the vacuum of the $\hat{A}_{j}$ modes,
that is, the state $|G\rangle$ of Eq. (\ref{Gpure}) we were seeking.
If the cooperativities are not large enough, the local decays will
introduce some entropy in the final state, so by tailoring them we
can even control the type of mixed Gaussian state $\hat{\rho}_{G}(\bar{\boldsymbol{n}})$
that we want to generate.

In the following we elaborate on these ideas and consider specific
examples.

\section{Effective Hamiltonian and limitations\label{Sec:FormalIdea}}

Let us move to the interaction picture defined by the transformation
operator
\begin{equation}
\hat{U}(t)=\exp\left(-\mathrm{i}\int_{0}^{t}dt'\hat{H}_{0}(t')\right),\;\text{with }\hat{H}_{0}=\sum_{j=1}^{N}\left\{ \omega_{j}\hat{n}_{j}+\left[\frac{\varepsilon_{j}}{2}+\sum_{m=1}^{2N}\Omega_{jm}\eta_{jm}\cos(\Omega_{jm}t+\phi_{jm})\right]\hat{\sigma}_{j}^{z}\right\} ,
\end{equation}
where states evolve according to the transformed Hamiltonian $\tilde{H}(t)=\hat{U}^{\dagger}(t)\hat{H}(t)\hat{U}(t)-\hat{H}_{0}(t)$,
which in turn takes the form
\begin{equation}
\tilde{H}(t)=\sum_{jl=1}^{N}g_{jl}\hat{\sigma}_{j}^{\dagger}\left[\alpha_{jl}(t)\hat{a}_{l}+\beta_{jl}(t)\hat{a}_{l}^{\dagger}\right]+\text{H.c.},\label{HintPic}
\end{equation}
with\begin{subequations}
\begin{align}
\alpha_{jl}(t) & =\sum_{n_{1}n_{2}...n_{2N}=-\infty}^{+\infty}J_{n_{1}}(2\eta_{j1})J_{n_{2}}(2\eta_{j2})...J_{n_{2N}}(2\eta_{j,2N})e^{-\mathrm{i}\left(\omega_{l}-\varepsilon_{j}-\sum_{m=1}^{2N}n_{m}\Omega_{jm}\right)t}e^{\mathrm{i}\sum_{m=1}^{2N}n_{m}\phi_{jm}},\\
\beta_{jl}(t) & =\sum_{n_{1}n_{2}...n_{2N}=-\infty}^{+\infty}J_{n_{1}}(2\eta_{j1})J_{n_{2}}(2\eta_{j2})...J_{n_{2N}}(2\eta_{j,2N})e^{\mathrm{i}\left(\omega_{l}+\varepsilon_{j}+\sum_{m=1}^{2N}n_{m}\Omega_{jm}\right)t}e^{\mathrm{i}\sum_{m=1}^{2N}n_{m}\phi_{jm}}.
\end{align}
\end{subequations}where $J_{n>0}(2\eta)\underset{\eta\ll\sqrt{n+1}}{\longrightarrow}\eta^{n}/n!$
are the Bessel functions, which satisfy $J_{-n}(2\eta)=(-1)^{n}J_{n}(2\eta)$.

Let us denote the oscillation frequencies of the different terms by\begin{subequations}
\begin{align}
\nu_{jl;\boldsymbol{n}}^{(\alpha)} & =\omega_{l}-\varepsilon_{j}-\sum_{m=1}^{2N}n_{m}\Omega_{jm}=\omega_{l}(1+n_{l}-n_{N+l})-\varepsilon_{j}\left(1+\sum_{m=1}^{2N}n_{m}\right)+\sum_{l\neq k=1}^{N}\omega_{k}(n_{k}-n_{N+k}),\label{etaAlpha}\\
\nu_{jl;\boldsymbol{n}}^{(\beta)} & =\omega_{l}+\varepsilon_{j}+\sum_{m=1}^{2N}n_{m}\Omega_{jm}=\omega_{l}(1-n_{l}+n_{N+l})+\varepsilon_{j}\left(1+\sum_{m=1}^{2N}n_{m}\right)-\sum_{l\neq k=1}^{N}\omega_{k}(n_{k}-n_{N+k}).\label{etaBeta}
\end{align}
\end{subequations}where we have introduced a vector $\boldsymbol{n}=(n_{1},n_{2},...,n_{2N})$
containing the Bessel indices and used (\ref{ModulationFreqs}). Let
us also define a quantity that we will call the $\eta$-order, $|n|=\sum_{m=1}^{2N}|n_{m}|$,
which for each term $J_{n_{1}}(2\eta_{j1})J_{n_{2}}(2\eta_{j2})...J_{n_{2N}}(2\eta_{j,2N})$
provides the order of the polynomial approximation in the small modulation
amplitudes $\eta_{jm}$. We will say that an index combination $\boldsymbol{n}$
is resonant when $\nu_{jl;\boldsymbol{n}}^{(\chi)}=0$, where $\chi$
can be either $\alpha$ or $\beta$. For each $\alpha_{jl}$ and $\beta_{jl}$
we already have a resonant term at $\eta$-order $|n|=1$, since\begin{subequations}\label{n1resonances}
\begin{align}
\nu_{jl;n_{1}=0,...,n_{l-1}=0,n_{l}=-1,n_{l+1}=0,...n_{2N}=0}^{(\alpha)} & =0,\\
\nu_{jl;n_{1}=0,...,n_{N+l-1}=0,n_{N+l}=-1,n_{N+l+1}=0,...n_{2N}=0}^{(\beta)} & =0.
\end{align}
\end{subequations}Ideally, we would like any other resonances to
appear only at large $\eta$-order $|n|$, so that their corresponding
contribution to the coupling is highly suppressed as $\eta^{|n|}$.
Lower $\eta$-order frequencies, on the other hand, should satisfy
$|\nu_{jl;\boldsymbol{n}}^{(\chi)}|\gg|g_{jl}J_{n_{1}}(2\eta_{j1})J_{n_{2}}(2\eta_{j2})...J_{n_{2N}}(2\eta_{j,2N})|$,
so that their contribution can be neglected by virtue of the rotating-wave
approximation. We later show that indeed this is not the case, and
we already have unavoidable resonances at $\eta$-order $|n|=3$.
For the sake of argumentation, let us however proceed for now assuming
that all contributions are negligible except the $|n|=1$ ones in
(\ref{n1resonances}), and we'll come back to this $|n|=3$ resonances
later. Under such assumption, the effective couplings can be rewritten
as\begin{subequations}
\begin{align}
\alpha_{jl} & \approx J_{0}(2\eta_{j1})...J_{0}(2\eta_{j,l-1})J_{-1}(2\eta_{jl})J_{0}(2\eta_{j,l+1})...J_{0}(2\eta_{j,2N})e^{-\mathrm{i}\phi_{jl}}\approx-\eta_{jl}e^{-\mathrm{i}\phi_{jl}},\\
\beta_{jl} & \approx J_{0}(2\eta_{j1})...J_{0}(2\eta_{j,N+l-1})J_{-1}(2\eta_{j,N+l})J_{0}(2\eta_{j,N+l+1})...J_{0}(2\eta_{j,2N})e^{-\mathrm{i}\phi_{j,N+l}}\approx-\eta_{j,N+l}e^{-\mathrm{i}\phi_{j,N+l}},
\end{align}
\end{subequations}where we assume that the modulation amplitudes
$\eta_{jl}$ are small enough such that the lowest-order approximation
of the Bessel functions hold. The interaction-picture Hamiltonian
(\ref{HintPic}) turns then into an effective Hamiltonian 
\begin{equation}
\tilde{H}_{\text{eff}}=-\sum_{j=1}^{N}\left[\sum_{l=1}^{N}g_{jl}\left(\eta_{jl}e^{-\mathrm{i}\phi_{jl}}\hat{a}_{l}+\eta_{j,N+l}e^{-\mathrm{i}\phi_{j,N+l}}\hat{a}_{l}^{\dagger}\right)\right]\hat{\sigma}_{j}^{\dagger}+\text{H.c.},
\end{equation}
which has exactly the form in (\ref{Heff}), making the correspondence\begin{subequations}
\begin{align}
\bar{g}_{j}\hat{A}_{j} & =\sum_{l=1}^{N}g_{jl}\left(\eta_{jl}e^{-\mathrm{i}\phi_{jl}}\hat{a}_{l}+\eta_{j,N+l}e^{-\mathrm{i}\phi_{j,N+l}}\hat{a}_{l}^{\dagger}\right)\\
 & \Downarrow\nonumber \\
\bar{g}_{j}\mathcal{A}_{jl}=g_{jl}\eta_{jl}e^{-\mathrm{i}\phi_{jl}}\; & ,\;\bar{g}_{j}\mathcal{B}_{jl}=g_{jl}\eta_{j,N+l}e^{-\mathrm{i}\phi_{j,N+l}},\;\;j,l=1,2,...,N.\label{MatchingEtaAB}
\end{align}
\end{subequations}Now, since the modulation amplitudes $\eta_{jm}$
and phases $\phi_{jm}$ can be freely chosen (just with the requirement
that the amplitudes must be small), this expression seems to suggest
that we indeed can access any multimode Gaussian state we want, just
with the subtlety that the effective couplings could become too small,
leading to slow cooling rates $|\bar{g}_{j}|^{2}/\gamma_{j}$. More
explicitly, let us consider the case of homogeneous coupling, $g_{jl}=g\,\forall jl$,
so that (\ref{MatchingEtaAB}) is recasted as
\begin{equation}
\frac{\bar{g}_{j}}{g}|\mathcal{A}_{jl}|=\eta_{jl}\;,\;\frac{\bar{g}_{j}}{g}|\mathcal{B}_{jl}|=\eta_{j,N+l}\;,\;\phi_{jl}=-\text{arg}\{\mathcal{A}_{jl}\}\;,\;\phi_{j,N+l}=-\text{arg}\{\mathcal{B}_{jl}\}.
\end{equation}
These expressions fix the phases $\phi_{jm}$. We can then make an
explicit choice for the amplitudes as well as follows. Assuming $\mathcal{A}_{j1}\neq0$
(but note that this is not a strong assumption, as the construction
we next make can be trivially adapted if this is not satisfied), we
start by fixing $\{\eta_{j1}\}_{j=1,2,...,N}$ to whatever value we
want. This fixes the rest of amplitudes as
\begin{equation}
\eta_{j,l}=\frac{|\mathcal{A}_{jl}|}{|\mathcal{A}_{j1}|}\eta_{j1},\;\eta_{j,N+l}=\frac{|\mathcal{B}_{jl}|}{|\mathcal{A}_{j1}|}\eta_{j1}.\label{EtaChoiceGeneralGaussian}
\end{equation}
The effective couplings, on the other hand, can be found from the
second condition in (\ref{GaussianUnitaryConditions}), whose diagonal
reads $\sum_{l=1}^{N}(|\mathcal{A}_{jl}|^{2}-|\mathcal{B}_{jl}|^{2})=1$
$\forall j$, which can be recasted as
\begin{equation}
\bar{g}_{j}^{2}=g^{2}\sum_{l=1}^{N}(\eta_{jl}^{2}-\eta_{j,N+l}^{2})=g^{2}\eta_{j1}^{2}\sum_{l=1}^{N}\frac{|\mathcal{A}_{jl}|^{2}-|\mathcal{B}_{jl}|^{2}}{|\mathcal{A}_{j1}|^{2}}.\label{gEFFgeneralGaussian}
\end{equation}
In general, Gaussian states are more entangled the larger the weights
of the $\hat{a}_{l}^{\dagger}$ terms are in $\hat{A}_{j}$. In turn,
this means that the more entanglement we want, the closer $\sum_{l=1}^{N}|\mathcal{A}_{jl}|^{2}$
and $\sum_{l=1}^{N}|\mathcal{B}_{jl}|^{2}$ will get, making the effective
couplings $\bar{g}_{j}$ smaller. Hence, we should prove through the
examples that it is possible to find a good balance between all these
features (scalable and large entanglement with reasonable cooling
rates).

Before moving on to examples, we need to comment about the limitations
imposed by the resonance at $\eta$-order $|n|=3$. In particular,
note that for a given frequency $\nu_{jl;\boldsymbol{n}}^{(\chi)}$
it is always possible to find $N-1$ resonances with $|n|=3$ that
make it vanish exactly: we can make each of the three terms adding
up in the final forms of (\ref{etaAlpha}) and (\ref{etaBeta}) vanish
independently by choosing $(n_{l},n_{N+l})=(0,1)$ for $\nu_{jl;\boldsymbol{n}}^{(\alpha)}$
and $(n_{l},n_{N+l})=(1,0)$ for $\nu_{jl;\boldsymbol{n}}^{(\beta)}$,
and then $n_{k}=n_{N+k}=-1$ for any other $k\neq l$. Hence, a more
precise expression for the couplings $\alpha_{jl}$ and $\beta_{jl}$
would be\begin{subequations}\label{3corrections}
\begin{align}
\alpha_{jl} & \approx-\eta_{jl}e^{-\mathrm{i}\phi_{jl}}+\sum_{l\neq k=1}^{N}\eta_{j,N+l}\eta_{jk}\eta_{j,N+k}e^{\mathrm{i}(\phi_{j,N+l}-\phi_{jk}-\phi_{j,N+k})},\\
\beta_{jl} & \approx-\eta_{j,N+l}e^{-\mathrm{i}\phi_{j,N+l}}+\sum_{l\neq k=1}^{N}\eta_{j,l}\eta_{jk}\eta_{j,N+k}e^{\mathrm{i}(\phi_{j,l}-\phi_{jk}-\phi_{j,N+k})},
\end{align}
\end{subequations}Assuming that all the amplitudes are of the same
order $\eta$, we then see that by neglecting this $|n|=3$ contribution,
we are making a relative mistake of order $N\eta^{2}$ in the worst
case. Depending on the accuracy with which we want to generate the
Gaussian state, this might need to be considered carefully. Of course,
one can always include this contribution when doing the matching (\ref{MatchingEtaAB})
and choose the amplitudes and phases accordingly, but then the construction
becomes more cumbersome. We will come back in the examples to the
limits that neglecting this $|n|=3$ contribution sets on the fidelity
of the final state, which we'll show not to be very strong even for
$N=10$ and large entanglement levels.

\section{Example: continuous-variable GHZ states\label{Sec:Example}}

\subsection{GHZ states\label{Sec:GHZ}}

As a specific example, we next consider the generation of continuous-variable
GHZ states of different number of modes $N$ \citep{Gaussian2,GHZ1,GHZ2}.
In the unphysical limit of perfect entanglement, these states converge
to the pure unnormalizable one
\begin{equation}
|\text{GHZ}_{N}\rangle=\int_{\mathbb{R}}dx\;\bigotimes_{j=1}^{N}|x\rangle,
\end{equation}
where $|x\rangle$ are the eigenstates of the position quadratures
$\hat{x}_{j}$. The most characteristic feature of these states is
that they show perfect correlation between all positions, as well
as a well-defined center-of-mass momentum, since $|\text{GHZ}_{N}\rangle$
is an eigenstate of the operators $\{\hat{x}_{j}-\hat{x}_{l}\}_{jl=1,...,N}$
and $\sum_{j=1}^{N}\hat{p}_{j}$ with zero eigenvalue. It is common
to summarize these correlations through the variances:
\begin{equation}
V\left(\frac{\hat{x}_{1}-\hat{x}_{2}}{\sqrt{2}}\right)=V\left(\frac{\hat{x}_{2}-\hat{x}_{3}}{\sqrt{2}}\right)=...=V\left(\frac{\hat{x}_{N-1}-\hat{x}_{N}}{\sqrt{2}}\right)=V\left(\frac{\hat{p}_{1}+\hat{p}_{2}+...+\hat{p}_{N}}{\sqrt{N}}\right)=0,\label{PerfectGHZcorrelation}
\end{equation}
where $V(\hat{B})=\langle\hat{B}^{2}\rangle-\langle\hat{B}\rangle^{2}$.
Note that tracing out any of the modes turns the state of the remaining
modes into the completely separable one $\int_{\mathbb{R}}dx\;\bigotimes_{j=1}^{N-1}|x\rangle\langle x|$,
showing that this is a state with genuine multipartite entanglement.
Note that for $N=2$, this is just the well known EPR or two-mode
perfectly-squeezed vacuum state \citep{CNBbook,Gaussian1,Gaussian2,ParisBook,CNB-QOnotes}.

While this state is unphysical, one can easily build a physical one
leading to the same physics \citep{Gaussian2,GHZ1,GHZ2}. For this,
we just relax the perfect-correlation condition (\ref{PerfectGHZcorrelation})
as\begin{subequations}\label{GHZcorrelations}
\begin{align}
V\left(\frac{\hat{p}_{1}+\hat{p}_{2}+...+\hat{p}_{N}}{\sqrt{N}}\right) & =e^{-2r_{1}},\\
V\left(\frac{\hat{x}_{1}-\hat{x}_{2}}{\sqrt{2}}\right) & =V\left(\frac{\hat{x}_{2}-\hat{x}_{3}}{\sqrt{2}}\right)=...=V\left(\frac{\hat{x}_{N-1}-\hat{x}_{N}}{\sqrt{2}}\right)=e^{-2r_{2}},
\end{align}
\end{subequations}for some finite real and positive parameters $r_{1}$
and $r_{2}$. Now the correlations are not perfect, but for large
$r_{n}$ they are still well beyond what's achievable for a coherent
state (corresponding to $r_{1}=r_{2}=0$) or mixtures of coherent
states, all these known as classical states, since they lead to a
positive and normalizable Glauber-Sudarshan distribution \citep{CarmichaelBook};
in other words, states satisfying (\ref{GHZcorrelations}) are non-classical
according to this notion of non-classicality. States satisfying (\ref{GHZcorrelations})
can be built in a very neat way by starting with $N$ single-mode
squeezed states ($N-1$ in position and 1 in momentum) and mixing
them in a succession of beam-splitters for neighboring modes \citep{Gaussian2,GHZ1,GHZ2}.
In particular, consider the following state
\begin{equation}
|\text{GHZ}_{N}(r_{1},r_{2})\rangle=\underbrace{\hat{B}_{N-1,N}(\theta_{N-1})...\hat{B}_{23}(\theta_{2})\hat{B}_{12}(\theta_{1})\hat{S}_{N}(-r_{2})...\hat{S}_{2}(-r_{2})\hat{S}_{1}(r_{1})}_{\hat{G}}|vac\rangle_{a},\label{GHZstate}
\end{equation}
with\begin{subequations}\label{GaussianUnitariesAction}
\begin{align}
\hat{S}_{j}(r) & =e^{\frac{r}{2}\left(\hat{a}_{j}^{\dagger2}-\hat{a}_{j}^{2}\right)}\;\Longrightarrow\;\hat{S}_{j}^{\dagger}(r)\hat{a}_{j}\hat{S}_{j}(r)=\hat{a}_{j}\cosh r+\hat{a}_{j}^{\dagger}\sinh r,\\
\hat{B}_{jl}(\theta) & =e^{\theta\left(\hat{a}_{j}\hat{a}_{l}^{\dagger}-\hat{a}_{j}^{\dagger}\hat{a}_{l}\right)}e^{\mathrm{i}\pi\hat{a}_{l}^{\dagger}\hat{a}_{l}}\;\Longrightarrow\;\left\{ \begin{array}{c}
\hat{B}_{jl}^{\dagger}(\theta)\hat{a}_{j}\hat{B}_{jl}(\theta)=\hat{a}_{j}\cos\theta+\hat{a}_{l}\sin\theta\\
\hat{B}_{jl}^{\dagger}(\theta)\hat{a}_{l}\hat{B}_{jl}(\theta)=-\hat{a}_{l}\cos\theta+\hat{a}_{j}\sin\theta
\end{array}\right.,
\end{align}
\end{subequations}and beam-splitter angles given by
\begin{equation}
\cos\theta_{n}=\frac{1}{\sqrt{N-n+1}},\:\:\sin\theta_{n}=\sqrt{\frac{N-n}{N-n+1}}.
\end{equation}
The GHZ state (\ref{GHZstate}) is already written as the action of
a Gaussian unitary $\hat{G}$ on the vacuum of the original modes
as in (\ref{Gpure}). Moreover, we know from (\ref{GaussianUnitariesAction})
how each of the unitaries act as a linear operation on the annihilation
and creation operators:
\begin{equation}
\hat{S}_{j}^{\dagger}(r)\hat{\boldsymbol{\alpha}}\hat{S}_{j}(r)=\mathcal{S}_{j}(r)\hat{\boldsymbol{\alpha}},\;\hspace{1cm}\hat{B}_{jl}^{\dagger}(\theta)\hat{\boldsymbol{\alpha}}\hat{B}_{jl}(\theta)=\mathcal{B}_{jl}(\theta)\hat{\boldsymbol{\alpha}},
\end{equation}
Here $\mathcal{S}_{j}(r)$ is a matrix equal to the $2N\times2N$
identity, except for entries $\cosh r$ at elements $(j,j)$ and $(N+j,N+j)$,
and entries $\sinh r$ at elements $(j,N+j)$ and $(N+j,j)$. On the
other hand, $\mathcal{B}_{jl}(\theta)$ is also a matrix equal to
the $2N\times2N$ identity, except for entries $\cos\theta$ at elements
$(j,j)$ and $(N+j,N+j)$, entries $-\cos\theta$ at elements $(l,l)$
and $(N+l,N+l)$, and entries $\sin\theta$ at elements $(j,l)$,
$(l,j)$, $(N+j,N+l)$, and $(N+l,N+j)$. Hence, combining the action
of all unitaries we find $\hat{G}^{\dagger}\hat{\boldsymbol{\alpha}}\hat{G}=\mathcal{G}\hat{\boldsymbol{\alpha}}$,
with a matrix
\begin{equation}
\mathcal{G}=\mathcal{B}_{N-1,N}(\theta_{N-1})...\mathcal{B}_{23}(\theta_{2})\mathcal{B}_{12}(\theta_{1})\mathcal{S}_{N}(-r_{2})...\mathcal{S}_{2}(-r_{2})\mathcal{S}_{1}(r_{1})\equiv\left(\begin{array}{cc}
\mathcal{A} & \mathcal{B}\\
\mathcal{B}^{*} & \mathcal{A}^{*}
\end{array}\right),
\end{equation}
whose upper-left and upper-right blocks correspond to the matrices
$\mathcal{A}$ and $\mathcal{B}$ that define the $\boldsymbol{\hat{A}}$
operators in (\ref{A}).

In order to check that this construction leads to the desired GHZ
state with correlations (\ref{GHZcorrelations}), we can evaluate
the covariance matrix of the state (\ref{GHZstate}) using (\ref{CovarianceMatrixGaussian})
with $\bar{\mathcal{N}}=0$, that is,
\begin{equation}
V_{\text{GHZ}}=\mathcal{T}\left(\begin{array}{cc}
\mathcal{A}\mathcal{B}^{T} & \mathcal{A}\mathcal{A}^{\dagger}\\
\mathcal{B}^{*}\mathcal{B}^{T} & \mathcal{B}^{*}\mathcal{A}^{\dagger}
\end{array}\right)\mathcal{T}^{T}-\mathrm{i}\Omega.\label{GHZcovarianceMatrix}
\end{equation}
One can easily check (better with the help of some symbolic program
to handle the matrix multiplications and diagonalization) that the
covariance matrix has eigenvalues $e^{\pm2r_{1}}$ and $e^{\pm2r_{2}}$,
the latter with $N-1$ degeneracy. The corresponding eigenvectors
are $(0,0,...,0,1,1,...,1)^{T}$ for $e^{-2r_{1}}$ and $\{(1,-1,0,...,0)^{T},(0,1,-1,0,...,0)^{T},...,(0,...,1,-1,0,...,0)\}$
for $e^{-2r_{2}}$, which correspond precise to the desired quadratures
when multiplied by $\hat{\boldsymbol{r}}$.

\subsection{Generation of GHZ states with our scheme\label{Sec:GHZidea}}

Let us now particularize to these GHZ states the choice of modulation
amplitudes $\eta_{jm}$ and phases $\phi_{jm}$ that we did in (\ref{EtaChoiceGeneralGaussian})
for general Gaussian states, and discuss what we find. Taking for
simplicity homogeneous couplings, $g_{jl}=g$, and using the matrices
$\mathcal{A}$ and $\mathcal{B}$ built as explained above for the
GHZ state, it is not difficult to find by inspection the following
forms for the amplitudes and phases for arbitrary number of modes
$N$\begin{subequations}\label{GHZmodulation}
\begin{align}
\eta_{jl} & =\eta_{j1}\times\left\{ \begin{array}{cc}
1, & l=1\\
0, & l>j+1\\
\sqrt{\frac{N(N-l+1)}{(N-l+2)}}\frac{\cosh r_{2}}{\cosh r_{1}}, & l=j+1\\
\sqrt{\frac{N}{(N-l+2)(N-l+1)}}\frac{\cosh r_{2}}{\cosh r_{1}}, & 1<l<j+1
\end{array}\right.,\\
\eta_{j,N+l} & =\eta_{jl}\times\left\{ \begin{array}{cc}
\tanh r_{1} & l=1\\
\tanh r_{2} & l>1
\end{array}\right.,\\
\phi_{jl} & =\left\{ \begin{array}{cc}
\pi, & l=1\text{ or }l>j\\
0 & \text{otherwise}
\end{array}\right.,\;\\
\phi_{j,N+l} & =\left\{ \begin{array}{cc}
\pi, & l=j+1\\
0 & \text{otherwise}
\end{array}\right.,
\end{align}
\end{subequations}where in these expressions $j,l=1,2,...,N$. There
are several notable things to mention here. First, assuming that we
take all $\eta_{j1}$ of the same order, say $\eta_{j1}=\eta\,\forall j$,
note that the largest amplitude is $\eta_{12}=\sqrt{N-1}\eta$ (further
assuming $r_{1}=r_{2}$ for simplicity), which we need to make sure
stays much smaller than 1, e.g., we need to choose $\eta=0.1/\sqrt{N-1}$.
On the other hand, inserting these expressions in (\ref{gEFFgeneralGaussian})
leads to the effective couplings
\begin{equation}
\bar{g}_{j}=\frac{\sqrt{N}\eta_{j1}}{\cosh r_{1}}g.
\end{equation}
Remarkably, the effective couplings do not depend on $r_{2}$, but
decrease exponentially with $r_{1}$. Also, they are dressed by a
factor $\sqrt{N}$, so they don't `feel' the $1/\sqrt{N-1}$ reduction
of $\eta$ mentioned above. In other words, the cooling rates are
approximately independent of $N$.

With these considerations, we see that superconducting-circuit parameters
similar to the ones we considered for the single-mode case \citep{CNB14}
would work just as well for the generation of multimode GHZ states.
The only potential issue we need to be careful with is choosing the
qubit and mode frequencies such that there are no multi-photon resonances.
Essentially, this just requires that all frequency differences are
large enough with respect to the couplings $|g|$. Taking (similarly
to our previous works \citep{Porras12,CNB14} and consistently with
experimental values \citep{JJGRbook,Blais21,Devoret13}) $g/2\pi=40$
MHz, $\varepsilon_{1}/2\pi=10$ GHz, and $\omega_{1}/2\pi=4.5$ GHz,
and assuming we take all other frequencies equally spaced as $\{\varepsilon_{j}=\varepsilon_{1}-10gj,\omega_{j}=\omega_{1}-10gj\}_{j=2,3,...,N}$,
we can reach $N=9$ while still keeping the lowest frequency above
$1$ GHz, which is reasonable for superconducting circuits. Of course,
way larger $N$ can be obtained by decreasing the spacing between
modes, and we have indeed checked that even taking $|\omega_{j}-\omega_{j+1}|=g$
we can still satisfy the conditions required for $\alpha_{jl}$ and
$\beta_{jl}$ to not receive extra higher $\eta$-order contributions.

\subsection{Limits imposed by the $|n|=3$ resonances\label{Sec:GHZlimits}}

With the GHZ example at hand, we can now give more quantitative details
about the error that one would make when not considering higher $\eta$-order
resonances, in particular those occurring at $|n|=3$. In order to
do this, we can simply compute the fidelity or overlap between the
GHZ states with and without the correction given in (\ref{3corrections}),
denoting by $\tilde{\mathcal{A}}$ and $\tilde{\mathcal{B}}$ the
matrices of the Gaussian state including the correction, as we explain
next.

The interaction-picture Hamiltonian (\ref{HintPic}) still has the
form of the effective Hamiltonian (\ref{Heff}), $\sum_{j=1}^{N}\tilde{g}_{j}\hat{\sigma}_{j}[\sum_{l=1}^{N}(\tilde{\mathcal{A}}_{jl}\hat{a}_{l}+\tilde{\mathcal{B}}_{jl}\hat{a}_{l}^{\dagger})]+\text{H.c.}$,
but now with the correspondences $\tilde{g}_{j}\tilde{\mathcal{A}}_{jl}=-g\alpha_{jl}$
and $\tilde{g}_{j}\tilde{\mathcal{B}}_{jl}=-g\beta_{jl}$, with $\alpha_{jl}$
and $\beta_{jl}$ given by (\ref{3corrections}). The qubits will
now cool the modes to the ground state of a Gaussian state with modified
matrices and effective couplings, as given by
\begin{equation}
\tilde{g}_{j}=g\sqrt{\sum_{k=1}^{N}(|\alpha_{jk}|^{2}-|\beta_{jk}|^{2})},\;\tilde{\mathcal{A}}_{jl}=-\frac{\alpha_{jl}}{\sqrt{\sum_{k=1}^{N}(|\alpha_{jk}|^{2}-|\beta_{jk}|^{2})}},\;\tilde{\mathcal{B}}_{jl}=-\frac{\beta_{jl}}{\sqrt{\sum_{k=1}^{N}(|\alpha_{jk}|^{2}-|\beta_{jk}|^{2})}},
\end{equation}
where the modulation amplitudes $\eta_{jm}$ and phases $\phi_{jm}$
are chosen as (\ref{GHZmodulation}) for the GHZ example. Given these
expressions, and using (\ref{CovarianceMatrixGaussian}), we can then
build the covariance matrix of the modified Gaussian state as 
\begin{equation}
\tilde{V}_{\text{GHZ}}=\mathcal{T}\left(\begin{array}{cc}
\tilde{\mathcal{A}}\tilde{\mathcal{B}}^{T} & \tilde{\mathcal{A}}\tilde{\mathcal{A}}^{\dagger}\\
\tilde{\mathcal{B}}^{*}\tilde{\mathcal{B}}^{T} & \tilde{\mathcal{B}}^{*}\tilde{\mathcal{A}}^{\dagger}
\end{array}\right)\mathcal{T}^{T}-\mathrm{i}\Omega.\label{GHZmodifiedCovarianceMatrix}
\end{equation}
Now all that is left is comparing the ideal Gaussian state $|\text{GHZ}_{N}\rangle$
with covariance matrix (\ref{GHZcovarianceMatrix}) and this modified
one, that we denote by $\widetilde{|\text{GHZ}_{N}\rangle}$. We denote
the corresponding Wigner functions by $W_{\text{GHZ}}(\boldsymbol{r})$
and $\tilde{W}_{\text{GHZ}}(\boldsymbol{r})$, which are Gaussians
of zero mean in both cases. Now, since they are pure states, we compare
them through the overlap, which is easily evaluated as \citep{CNBbook,Gaussian1,Gaussian2,ParisBook,CNB-QOnotes}
\begin{align}
\left|\langle\text{GHZ}_{N}\widetilde{|\text{GHZ}_{N}\rangle}\right|^{2} & =(4\pi)^{N}\int_{\mathbb{R}^{2N}}d^{2N}\boldsymbol{r}W_{\text{GHZ}}(\boldsymbol{r})\tilde{W}_{\text{GHZ}}(\boldsymbol{r})\\
 & =\frac{(4\pi)^{N}}{(2\pi)^{2N}\sqrt{\det\{V_{\text{GHZ}}\}\det\{\tilde{V}_{\text{GHZ}}\}}}\int_{\mathbb{R}^{2N}}d^{2N}\boldsymbol{r}e^{-\frac{1}{2}\boldsymbol{r}^{T}(V_{\text{GHZ}}^{-1}+\tilde{V}_{\text{GHZ}}^{-1})\boldsymbol{r}}=\frac{2^{N}}{\sqrt{\det\{V_{\text{GHZ}}^{-1}+\tilde{V}_{\text{GHZ}}^{-1}\}}},\nonumber 
\end{align}
where we have used $\det\{V_{\text{GHZ}}\}=1=\det\{\tilde{V}_{\text{GHZ}}\}$
since the states are pure. Using this expression and setting $\eta_{1j}=0.1/\sqrt{N-1}$,
for 90\% squeezing ($e^{-2r_{1}}=e^{-2r_{2}}=0.1$) we have checked
that the fidelity remains above 0.998 for as large $N$ as we have
been patient enough to compute ($N=10$). In fact, for $N=10$, we
have seen that the fidelity falls below 0.99, 0.95, and 0.9 only if
the squeezing exceeds, respectively, 95.3\%, 97.7\%, and 98.4\% ($e^{-2r_{1}}=e^{-2r_{2}}=0.047,$
$0.023$, and $0.016$). As for the effective couplings $\tilde{g}_{j}$,
we have checked that for any value of the squeezing, they are extremely
close to the original ones $\bar{g}_{j}$ (say, within 5 significant
digits).

In summary, for our purposes, the $|n|=3$ multi-photon resonances
do not seem to be a problem.

\section{Avoiding all-to-all coupling\label{Sec:AvoidingALLtoALL}}

Perhaps the main experimental hurdle of our proposal is the fact that
the model we present has connections of all modes to all qubits, which
is pretty unrealistic through direct coupling in current architectures.
Fortunately, effective ways such as those relying on resonator networks
can help \citep{Olivares17}. As a proof of concept, we consider here
a simpler situation: we next show that a chain of nearest-neighbor-coupled
modes, with each mode locally coupled to a single qubit, achieves
the type of Hamiltonian we need in the normal-mode basis of the chain.
We compare two types of chains that allow us for analytic calculations:
one with open boundaries and one with closed boundaries.

\subsection{Open bosonic chain}

Consider first the model
\begin{equation}
\hat{H}=\sum_{j=1}^{N-1}\left(\omega\hat{a}_{j}^{\dagger}\hat{a}_{j}-J\hat{a}_{j}\hat{a}_{j+1}^{\dagger}-J\hat{a}_{j}^{\dagger}\hat{a}_{j+1}+\frac{\varepsilon_{j}}{2}\hat{\sigma}_{j}^{z}\right)+\sum_{j=1}^{N}g_{j}(\hat{\sigma}_{j}+\hat{\sigma}_{j}^{\dagger})(\hat{a}_{j}+\hat{a}_{j}^{\dagger}).\label{Hchain}
\end{equation}
Note that we have taken all the mode frequencies and hoppings equal
(homogeneous chain) in order to be able to perform analytic calculations.
We can move to a normal-mode basis that diagonalizes the bosonic part
of this model, which we write as $\hat{\boldsymbol{a}}^{\dagger}\mathcal{M}\hat{\boldsymbol{a}}$,
with a tridiagonal matrix
\begin{equation}
\mathcal{M}=\left(\begin{array}{cccccc}
\omega & -J\\
-J & \omega & -J\\
 & -J & \omega & -J\\
 &  & \ddots & \ddots & \ddots\\
 &  &  & -J & \omega & -J\\
 &  &  &  & -J & \omega
\end{array}\right).
\end{equation}
This can be diagonalized \citep{Noschese13} as $\mathcal{M}=\mathcal{SDS}$,
where $\mathcal{D}=\text{diag}(\Delta_{1},\Delta_{2},...,\Delta_{N})$
is a diagonal matrix containing the eigenvalues $\Delta_{k}=\omega-2J\cos\left(\frac{k\pi}{N+1}\right)$
and $\mathcal{S}$ is a symmetric orthogonal matrix ($\mathcal{S}^{T}=\mathcal{S}$
and $\mathcal{S}^{T}\mathcal{S}=\mathcal{I}$) with elements $\mathcal{S}_{jk}=\sqrt{\frac{2}{N+1}}\sin\left(\frac{jk\pi}{N+1}\right)$.
In terms of the transformed bosonic operators $\hat{\boldsymbol{c}}=\mathcal{S}\hat{\boldsymbol{a}}$,
the bosonic part of the Hamiltonian takes then the form $\hat{\boldsymbol{c}}^{\dagger}\mathcal{D}\hat{\boldsymbol{c}}$,
so the total Hamiltonian is rewritten as
\[
\hat{H}=\sum_{k=1}^{N}\Delta_{k}\hat{c}_{k}^{\dagger}\hat{c}_{k}+\sum_{j=1}^{N}\frac{\varepsilon_{j}}{2}\hat{\sigma}_{j}^{z}+\sum_{jk=1}^{N}g_{jk}(\hat{\sigma}_{j}+\hat{\sigma}_{j}^{\dagger})(\hat{c}_{k}+\hat{c}_{k}^{\dagger}),
\]
where we have defined the couplings $g_{jk}=\sqrt{\frac{2}{N+1}}g_{j}\sin\left(\frac{jk\pi}{N+1}\right)$.
Note that the couplings are now reduced by a $\sqrt{N+1}$ factor,
but we will take these renormalized couplings as the ones whose magnitude
we fix, e.g., to 40 MHz as in the example of the previous section.
In addition, note that the couplings are modulated by a sinusoidal
function $\sin\left(\frac{jk\pi}{N+1}\right)$, which can vanish when
$jk=N+1$; one then needs to be careful to pick a prime $N+1$ in
this open chain configuration.

All nuances apart, we then see that in this normal-mode basis we obtain
the model we were seeking, with all modes coupled to all qubits. Note
that as $N$ increases, the difference between $\Delta_{k}$ and $\Delta_{k\pm1}$
decreases; we can estimate the worst case situation by considering
the mode $k$ at the top of the dispersion relation ($k=1$) and its
neighbor ($k=2$), whose difference is given by $2J[\cos\left(\frac{2\pi}{N+1}\right)-\cos\left(\frac{\pi}{N+1}\right)]\approx3\pi^{2}J/N^{2}$
for large $N$, showing that if we want to keep this difference on
the order of the largest coupling $\max(|g_{jk}|)\equiv g$, the hopping
will have to scale as the square of the number of modes, that is,
$J\geq gN^{2}/3\pi^{2}$. But also, note that the mode frequencies
spanning over an interval $4J$ around $\omega$ can't get close to
the qubit frequencies. For example, assuming that the qubit frequencies
are larger than the mode frequencies, we demand that the largest mode
frequency is much smaller than the smallest qubit frequency. Take
then $\varepsilon_{1}=2\pi\times10$ GHz, and the rest of qubit frequencies
spaced by $g$, so that the smallest one is $\varepsilon_{N}=2\pi\times10\text{ GHz}-(N-1)g$,
and assume that the smallest mode frequency $\omega_{\text{min}}\approx\omega-2J$
is equal to $2\pi\times$1 GHz, so that the largest must obbey $\omega_{\text{max}}\leq2\pi\times1\text{ GHz}+4J\sim2\pi\times1\text{ GHz}+4gN^{2}/3\pi^{2}$.
Taking $g=2\pi\times40$ MHz, we then get that $\varepsilon_{N}-\omega_{\text{max}}>35g$
as long as $N\leq35$, which is a huge number of modes. This proves
that the idea of working with a chain is feasible.

One more thing to consider is to which state we need to cool down
the normal modes $\hat{\boldsymbol{c}}$ in order to obtain a desired
Gaussian state of the original modes $\hat{\boldsymbol{a}}$. For
this, just keep in mind that the relation between these modes can
be written as $\hat{\boldsymbol{a}}=\mathcal{S}\hat{\boldsymbol{c}}$.
Then, applying $\mathcal{S}$ on (\ref{A}), we obtain the action
of the Gaussian unitary $\hat{G}$ that defines the target state $|G\rangle=\hat{G}|vac\rangle_{a}$
on the normal modes (note that $|vac\rangle_{a}$ is also the vacuum
state of the normal modes, $\hat{\boldsymbol{c}}|vac\rangle_{a}=0$,
since the transformation $\mathcal{S}$ is passive \citep{CNBbook}),
which define the new set of bosonic operators $\hat{\boldsymbol{C}}$
that we will need to cool down:
\begin{equation}
\hat{G}^{\dagger}\hat{\boldsymbol{c}}\hat{G}=\underbrace{\mathcal{S}\mathcal{A}\mathcal{S}}_{\mathcal{\mathcal{A}}^{(c)}}\hat{\boldsymbol{c}}+\underbrace{\mathcal{S}\mathcal{B}\mathcal{S}}_{\mathcal{\mathcal{B}}^{(c)}}\hat{\boldsymbol{c}}^{\dagger T}\equiv\boldsymbol{\hat{C}}.
\end{equation}
The qubit modulation
\begin{equation}
\sum_{j=1}^{N}\left[\sum_{k=1}^{N}\Omega_{jk}\eta_{jk}\cos(\Omega_{jk}t+\phi_{jk})\right]\hat{\sigma}_{j}^{z}
\end{equation}
 would now induce the effective Hamiltonian
\begin{equation}
\hat{H}_{\text{eff}}=-\sum_{j=1}^{N}\bar{g}_{j}\hat{C}_{j}\hat{\sigma}_{j}^{\dagger}+\text{H.c.},
\end{equation}
 as long as the modulation amplitudes and phases are chosen to satisfy
\begin{equation}
\bar{g}_{j}\mathcal{A}_{jk}^{(c)}=g_{jk}\eta_{jk}e^{-\mathrm{i}\phi_{jk}}\;,\;\bar{g}_{j}\mathcal{B}_{jk}^{(c)}=g_{jk}\eta_{j,N+k}e^{-\mathrm{i}\phi_{j,N+k}},
\end{equation}
where the only difference with the all-to-all connected model is that
the target matrices are $\mathcal{S}$-transformed, $\mathcal{A}^{(c)}$
and $\mathcal{B}^{(c)}$, and the modulation phases need to cancel
additional phases (signs) coming from some of the $\sin\left(\frac{kj\pi}{N+1}\right)$
terms in the couplings $g_{jk}$. Specifically, assuming couplings
$g_{jk}=g\sin\left(\frac{kj\pi}{N+1}\right)$ of equal magnitude except
for the sinusoidal modulation, we can make the same choices as we
did in previous sections:\begin{subequations}
\begin{align}
\eta_{jk} & =\left|\frac{\mathcal{A}_{jk}^{(c)}\sin\left(\frac{j\pi}{N+1}\right)}{\mathcal{A}_{j1}^{(c)}\sin\left(\frac{kj\pi}{N+1}\right)}\right|\eta_{j1},\\
\eta_{j,N+k} & =\left|\frac{\mathcal{B}_{jk}^{(c)}\sin\left(\frac{j\pi}{N+1}\right)}{\mathcal{A}_{j1}^{(c)}\sin\left(\frac{kj\pi}{N+1}\right)}\right|\eta_{j1},\\
\phi_{jk} & =\text{arg}\left\{ \sin\left(\frac{kj\pi}{N+1}\right)\right\} -\text{arg}\{\mathcal{A}_{jk}^{(c)}\},\\
\phi_{j,N+k} & =\text{arg}\left\{ \sin\left(\frac{kj\pi}{N+1}\right)\right\} -\text{arg}\{\mathcal{B}_{jk}^{(c)}\},
\end{align}
\end{subequations}where $\{\eta_{j1}\}_{j=1,2,...,N}$ are fixed
to whatever value we want, and the effective couplings read
\begin{equation}
\bar{g}_{j}^{2}=g^{2}\eta_{j1}^{2}\sum_{k=1}^{N}\frac{|\mathcal{A}_{jk}^{(c)}|^{2}-|\mathcal{B}_{jk}^{(c)}|^{2}}{|\mathcal{A}_{j1}^{(c)}|^{2}}\left|\frac{\sin\left(\frac{j\pi}{N+1}\right)}{\sin\left(\frac{kj\pi}{N+1}\right)}\right|.\label{gEFFgeneralGaussian-1}
\end{equation}

\subsection{Closed bosonic chain}

Consider next the model
\begin{equation}
\hat{H}=\sum_{j=1}^{N}\left(\omega\hat{a}_{j}^{\dagger}\hat{a}_{j}-Je^{\mathrm{i}\phi}\hat{a}_{j}\hat{a}_{j+1}^{\dagger}-Je^{-\mathrm{i}\phi}\hat{a}_{j}^{\dagger}\hat{a}_{j+1}+\frac{\varepsilon_{j}}{2}\hat{\sigma}_{j}^{z}\right)+\sum_{j=1}^{N}g_{j}(\hat{\sigma}_{j}+\hat{\sigma}_{j}^{\dagger})(\hat{a}_{j}+\hat{a}_{j}^{\dagger}).\label{Hchain-1}
\end{equation}
where again we take all the mode frequencies and hoppings equal (homogeneous
chain) in order to be able to perform analytic calculations. We will
see in a second that this forces us to introduce complex hoppings
$\phi\neq0$ (sometimes referred to as `external artificial Gauge
field') for our ideas to work, but real hoppings would work as well
as long as the chain is sufficiently inhomogeneous. Periodic boundaries
are assumed this time, that is, $\hat{a}_{N+1}\equiv\hat{a}_{1}$.
Let's move to the Fourier basis that diagonalizes the bosonic part
of this model
\begin{equation}
\hat{a}_{j}=\frac{1}{\sqrt{N}}\sum_{k=k_{\text{min}}}^{k_{\text{min}}+N-1}e^{2\pi\mathrm{i}jk/N}\hat{c}_{k}\;\Longleftrightarrow\;\hat{c}_{k}=\frac{1}{\sqrt{N}}\sum_{j=1}^{N}e^{-2\pi\mathrm{i}jk/N}\hat{a}_{j},
\end{equation}
where we choose to work in the first Brillouin zone so that $k_{\text{min}}=-N/2$
or $-(N-1)/2$ for even or odd $N$, respectively. Inserting this
expression in (\ref{Hchain-1}) and using the completeness relation
$\sum_{j=1}^{N}e^{2\pi\mathrm{i}j(k-k')/N}=N\delta_{k,k'}$, we easily
obtain
\begin{equation}
\hat{H}=\sum_{k=k_{\text{min}}}^{k_{\text{min}}+N-1}\Delta_{k}\hat{c}_{k}^{\dagger}\hat{c}_{k}+\sum_{j=1}^{N}\frac{\varepsilon_{j}}{2}\hat{\sigma}_{j}^{z}+\sum_{j=1}^{N}\sum_{k=k_{\text{min}}}^{k_{\text{min}}+N-1}(\hat{\sigma}_{j}+\hat{\sigma}_{j}^{\dagger})(g_{jk}\hat{c}_{k}+g_{jk}^{*}\hat{c}_{k}^{\dagger}),
\end{equation}
where we have defined the dispersion relation $\Delta_{k}=\omega-2J\cos(2\pi k/N-\phi)$
and the complex couplings $g_{jk}=e^{2\pi\mathrm{i}jk/N}g_{j}/\sqrt{N}$.
Note that the couplings are now reduced by a $\sqrt{N}$ factor, but
do not posses the sinusoidal modulation present in the open chain,
so in this case there is no restriction on the values of $N$. Note
that the condition that all mode frequencies must be different, $\Delta_{k}\neq\Delta_{k'\neq k}$,
imposes that $\phi$ cannot take certain values such as 0 or $\pi$,
for which $\Delta_{k}=\Delta_{-k}$. Hence, having complex hopping
in the homogeneous chain is a necessary condition. Of course, it might
be experimentally easier to work with an inhomogeneous chain (e.g.,
bosonic modes of unequal frequencies) and keep the hoppings real.
Also, note that as $N$ increases, the difference between $\Delta_{k}$
and $\Delta_{k\pm1}$ decreases; we can estimate the worst case situation
in the same way as we did above for the open chain, by considering
in this case a mode $k$ at the bottom of the dispersion relation
and a neighboring one, whose difference is given by $2J[1-\cos(2\pi/N)]\approx8\pi^{2}J/N^{2}$
for large $N$, showing that if we want to keep this difference on
the order of the largest coupling $\max(|g_{j}|/\sqrt{N})\equiv g$,
the hopping will have to scale again as the square of the number of
modes, that is, $J\geq gN^{2}/8\pi^{2}$. Taking $g=2\pi\times40$
MHz, we apply again the condition that the smallest qubit frequency
$\varepsilon_{N}=2\pi\times10\text{ GHz}-(N-1)g$ must be larger than
the largest Fourier-mode frequency $\omega_{\text{max}}\leq2\pi\times1\text{ GHz}+4J\sim2\pi\times1\text{ GHz}+gN^{2}/2\pi^{2}$,
obtaining $\varepsilon_{N}-\omega_{\text{max}}>50g$ as long as $N<50$,
which is again a huge number of modes. 

We finally need to consider again the state to which we need to cool
down the Fourier modes in order to obtain a desired Gaussian state
of the original modes. The relation between these modes can be written
now as $\hat{\boldsymbol{a}}=\mathcal{F}\hat{\boldsymbol{c}}$, with
$\mathcal{F}$ a unitary matrix with elements $\mathcal{F}_{jk}=e^{2\pi\mathrm{i}jk/N}/\sqrt{N}$
and $\hat{\boldsymbol{c}}=(\hat{c}_{k_{\text{min}}},\hat{c}_{k_{\text{min}+1}},...,\hat{c}_{k_{\text{min}}+N-1})^{T}$
collecting all the Fourier annihilation operators. Then, applying
$\mathcal{F}^{\dagger}$ on (\ref{A}), we obtain the action of the
Gaussian unitary $\hat{G}$ that defines the target state $|G\rangle=\hat{G}|vac\rangle_{a}$
on the Fourier modes, which defines the new set of bosonic operators
$\hat{\boldsymbol{C}}$ that we will need to cool down:
\begin{equation}
\hat{G}^{\dagger}\hat{\boldsymbol{c}}\hat{G}=\underbrace{\mathcal{F}^{\dagger}\mathcal{A}\mathcal{F}}_{\mathcal{\mathcal{A}}^{(c)}}\hat{\boldsymbol{c}}+\underbrace{\mathcal{F}^{\dagger}\mathcal{B}\mathcal{F}^{*}}_{\mathcal{\mathcal{B}}^{(c)}}\hat{\boldsymbol{c}}^{\dagger T}\equiv\boldsymbol{\hat{C}}.
\end{equation}
The qubit modulation
\begin{equation}
\sum_{j=1}^{N}\left[\sum_{k=k_{\text{min}}}^{k_{\text{min}}+2N-1}\Omega_{jk}\eta_{jk}\cos(\Omega_{jk}t+\phi_{jk})\right]\hat{\sigma}_{j}^{z}
\end{equation}
 would now induce the effective Hamiltonian
\begin{equation}
\hat{H}_{\text{eff}}=-\sum_{j=1}^{N}\bar{g}_{j}\hat{C}_{j}\hat{\sigma}_{j}^{\dagger}+\text{H.c.},
\end{equation}
 as long as the modulation amplitudes and phases are chosen to satisfy
\begin{equation}
\bar{g}_{j}\mathcal{A}_{jk}^{(c)}=g_{jk}\eta_{jk}e^{-\mathrm{i}\phi_{jk}}\;,\;\bar{g}_{j}\mathcal{B}_{jk}^{(c)}=g_{jk}^{*}\eta_{j,N+k}e^{-\mathrm{i}\phi_{j,N+k}}.
\end{equation}
Assuming couplings $g_{jk}=g\exp(2\pi\mathrm{i}jk/N)$ of equal magnitude,
we can make the same choices as we did in previous sections:\begin{subequations}
\begin{align}
\eta_{jk} & =\frac{|\mathcal{A}_{jk}^{(c)}|}{|\mathcal{A}_{jk_{\text{min}}}^{(c)}|}\eta_{jk_{\text{min}}},\;\eta_{j,N+k}=\frac{|\mathcal{B}_{jk}^{(c)}|}{|\mathcal{A}_{jk_{\text{min}}}^{(c)}|}\eta_{jk_{\text{min}}},\\
\phi_{jk} & =2\pi\frac{jk}{N}-\text{arg}\{\mathcal{A}_{jk}^{(c)}\}\;,\;\phi_{j,N+k}=-2\pi\frac{jk}{N}-\text{arg}\{\mathcal{B}_{jk}^{(c)}\},
\end{align}
\end{subequations}where $\{\eta_{jk_{\text{min}}}\}_{j=1,2,...,N}$
are fixed to whatever value we want as usual, and the effective couplings
read
\begin{equation}
\bar{g}_{j}^{2}=g^{2}\eta_{jk_{\text{min}}}^{2}\sum_{k=k_{\text{min}}}^{k_{\text{min}}+2N-1}\frac{|\mathcal{A}_{jk}^{(c)}|^{2}-|\mathcal{B}_{jk}^{(c)}|^{2}}{|\mathcal{A}_{jk_{\text{min}}}^{(c)}|^{2}}.\label{gEFFgeneralGaussian-1-1}
\end{equation}

\section{Concluding remarks\label{Sec:Conclusions}}

We have shown how to modulate a collection of qubits coupled to a
collection of bosonic modes so as to dissipatively steer the latter
into any desired multimode Gaussian state. All-to-all couplings can
be avoided by, for example, starting from a bosonic chain. We have
shown that our ideas are feasible through the example of GHZ states.
While we haven't presented it here owed to space limitations, we remark
that we have also worked out a completely different set of examples
consisting of various types of cluster states \citep{Cluster1,Cluster2,Cluster3,Cluster4},
finding no additional hurdles to those we have thoroughly discussed
already for GHZ states.

Note that once we know how to generate arbitrary Gaussian states through
cooling, we can also do it via lasing, just adapting the ideas we
presented in \citep{CNB14}. In particular, one would need to add
an auxiliary set of qubits, but modulated in such a way that the anti-Jaynes-Cummings
equivalent of (\ref{Heff}) is generated, that is, an effective Hamiltonian
of the form $-\sum_{j=1}^{N}\bar{g}_{j}(\hat{A}_{j}\hat{\sigma}_{j}+\hat{A}_{j}^{\dagger}\hat{\sigma}_{j}^{\dagger})$.
In order to accomplish this, one just needs to replace the correspondence
(\ref{MatchingEtaAB}) by \begin{subequations}
\begin{align}
\bar{g}_{j}\hat{A}_{j}^{\dagger} & =\sum_{l=1}^{N}g_{jl}\left(\eta_{jl}e^{-\mathrm{i}\phi_{jl}}\hat{a}_{l}+\eta_{j,N+l}e^{-\mathrm{i}\phi_{j,N+l}}\hat{a}_{l}^{\dagger}\right)\\
 & \Downarrow\nonumber \\
\bar{g}_{j}\mathcal{B}_{jl}^{*}=g_{jl}\eta_{jl}e^{-\mathrm{i}\phi_{jl}}\; & ,\;\bar{g}_{j}\mathcal{A}_{jl}^{*}=g_{jl}\eta_{j,N+l}e^{-\mathrm{i}\phi_{j,N+l}},\;\;j,l=1,2,...,N.\label{MatchingEtaAB-lasing}
\end{align}
\end{subequations}After tracing out the qubits used for cooling,
one is then left with the following master equation for the state
$\hat{\rho}$ of the bosonic modes and the auxiliary qubits \citep{CNB14}:
\begin{equation}
\frac{d\hat{\rho}}{dt}=\sum_{j=1}^{N}\left(\mathrm{i}\bar{g}_{j}\left[\hat{A}_{j}\hat{\sigma}_{j}+\hat{A}_{j}^{\dagger}\hat{\sigma}_{j}^{\dagger},\hat{\rho}\right]+\mathcal{D}_{A_{j}}[\hat{\rho}]+\mathcal{D}_{\sigma_{j}}[\hat{\rho}]\right).
\end{equation}
This is equivalent to the master equation of a collection of independent
single-qubit lasers \citep{CNB14} for each bosonic mode $\hat{A}_{j}$
(if you are not convinced, note that exchanging the dummy labels between
the qubit states, $|g\rangle_{j}\rightleftharpoons|e\rangle_{j}$,
is equivalent to a $\hat{\sigma}_{j}\rightleftharpoons\hat{\sigma}_{j}^{\dagger}$
swap, turning the qubit dissipation into pumping, and the Hamiltonian
into a Jaynes-Cummings one). This opens the possibility of experimentally
engineering nonclassical multimode lasing, something that sounds quite
exotic specially in the context of optical settings.
\begin{acknowledgments}
CNB thanks Valentina Hopekin for inspiration with the manuscript,
and acknowledges sponsorship from the Yangyang Development Fund, as
well as support from a Shanghai talent program and from the Shanghai
Municipal Science and Technology Major Project (Grant No. 2019SHZDZX01).
DP an JJGR acknowledge financial support from the Proyecto Sinérgico
CAM 2020 Y2020/TCS-6545 (NanoQuCoCM), the CSIC Interdisciplinary Thematic
Platform (PTI+) on Quantum Technologies (PTI-QTEP+) and from Spanish
project PID2021-127968NB-I00 (MCIU/AEI/FEDER, EU).
\end{acknowledgments}

\bibliographystyle{apsrev4-1}
\bibliography{CoolingToGaussianCircuitQED}

\end{document}